\documentclass[twocolumn,showpacs,amsmath,amssymb]{revtex4}
\usepackage[dvips]{graphicx}% Include figure files
\usepackage{dcolumn}% Align table columns on decimal point
\usepackage{bm}% bold math

\begin{document}

\preprint{}

\title{Roton energy gap and spontaneous symmetry breaking}% Forced % line breaks with \\

\author{Junpei Harada}
 \email[]{E-mail:harada@sci.niihama-nct.ac.jp}
 \affiliation{Niihama National College of Technology, Niihama, 792-8580, Japan}%
\date{November 9, 2009}% It is always \today, today,

\begin{abstract}
We study elementary excitations in superfluid helium-4 employing an approach based on the spontaneous symmetry breaking. In particular, we calculate the roton energy gap at zero temperature $\Delta (0)$. The relation that we have derived is $\Delta (0) / k_B T_\lambda = 4$. The theoretical value of $\Delta (0)/k_B$ is 8.707 K, which is significantly close to the experimental value of 8.712 K. The deviation between the theoretical and experimental values is less than 0.1\%. 
\end{abstract}

\pacs{67.25.D-, 67.25.dj, 67.25.dt}% PACS, the Physics and Astronomy
%\keywords{keywords}%Use showkeys class option if keyword

\maketitle

\section{Introduction}
Recently, the author has derived the following formula for the $\lambda$-temperature in superfluid helium-4~\cite{Harada:2007}
\begin{eqnarray}
 T_\lambda = \frac{\rho_0 \pi \hbar^2 \sigma}{m^2 k_B}, \label{eq:lambda}
\end{eqnarray}
where $\rho_0$ is the density of liquid helium-4 at zero temperature, $m$ and $\sigma$ are the mass and the hard sphere diameter of a helium atom, respectively. Thus, $\lambda$-temperature is determined only by the {\it three} physical parameters, $\rho_0, m$ and $\sigma$. Among these parameters, the hard sphere diameter $\sigma$ is especially important, because it is a crucial parameter for hard-core repulsive interactions at short distances. Using eq.~(\ref{eq:lambda}), one can obtain the theoretical value of $T_\lambda$, because the experimental values of $\rho_0$, $m$ and $\sigma$ are known. Eq.~(\ref{eq:lambda}) gives $T_\lambda = 2.194$ K~\cite{Harada:2007}, which is very close to the experimental value of 2.1768 K~\cite{Donnelly:1998}. The deviation between the theoretical and experimental values is less than 1$\%$. In this meaning, our approach succeeded. However, in the previous paper~\cite{Harada:2007}, we did not study the elementary excitations.

Although the roton, which is an elementary excitation with finite energy gap $\Delta$, has been extensively studied in the past, its physical nature still remains unclear~\cite{Donnelly:1997}. In this paper, we calculate the roton energy gap at zero temperature $\Delta (0)$ to understand the physical nature of roton. First, we calculate the roton energy gap using the effective potential of the mean field effective field theory. Second, we introduce the curvature matrix which allows us to calculate the energy gap directly. The relation that we have derived is very simple, $\Delta (0)/k_B T_\lambda = 4$. This relation gives $\Delta (0)/k_B = 8.707$ K, which is very close to the experimental value of 8.712 K. Finally, we comment on an open problem about the sound speed of phonon. In the next section, we begin with a brief review of ref.~\cite{Harada:2007} for self-consistency of this paper. 

\section{Spontaneous symmetry breaking and $\lambda$-temperature}
The superfluid phase of liquid helium-4 is described by the Gross-Pitaevskii (GP) effective field theory, in which the cooperative state of many interacting helium atoms is represented with a single complex scalar field $\varphi$. GP theory is defined by  
\begin{eqnarray}
 {\cal L} & = & 
                 i\hbar \varphi^* \frac{\partial}{\partial t}\varphi
               +\frac{\hbar^2}{2m} \varphi^* \nabla^2 \varphi
               +\mu \varphi^* \varphi 
               - \frac{\lambda}{2}\left(\varphi^* \varphi \right)^2, \label{eq:GP}
\end{eqnarray}
where $\varphi$ is a one-component complex scalar field with both amplitude and phase,
$m$ is the mass of a helium atom, $\mu$ is the chemical potential, 
$\lambda = 2\pi\hbar^2\sigma/m$ is the coupling constant for hard-core repulsive interactions at short distances and $\sigma$ is the hard sphere diameter of a helium atom, respectively. Thus, GP theory is a nonrelativistic version of the Goldstone model in particle physics. GP Lagrangian~(\ref{eq:GP}) is invariant under the {\it global} $U(1)$ transformations; $\varphi \rightarrow e^{i\theta}\varphi$, $\varphi^* \rightarrow e^{-i\theta}\varphi^*$ with $\partial \theta / \partial t = \nabla \theta = 0$. Because $U(1)$ symmetry is not gauged, no vector gauge field appears.

\begin{figure}[t]
\includegraphics[width=5cm]{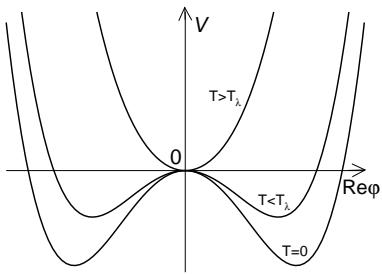}
\caption{\label{fig:transition} The schematic temperature dependence of the effective potential $V$. $\varphi = 0$ is not stable for $T < T_\lambda$.}
\end{figure}

In the GP theory, the $\lambda$-transition is described as the spontaneous symmetry breaking of the $U(1)$ symmetry. One can understand this statement as follows. The effective potential $V$ at the tree-level approximation is given by the third and the fourth terms in eq.~(\ref{eq:GP}), which does not contain the derivatives; 
\begin{eqnarray}
	V = -\mu\varphi^*\varphi + \frac{\lambda}{2}(\varphi^*\varphi)^2. \label{eq:potential}
\end{eqnarray}
As shown in ref.~\cite{Harada:2007}, the $\lambda$-temperature $T_\lambda$ is determined only by this potential. Here it should be noted that the chemical potential $\mu$ becomes {\it positive} below the $\lambda$-temperature~\cite{Harada:2007}, and then the ground states become degenerate for $T < T_\lambda$. Fig.~\ref{fig:transition} shows the schematic temperature dependence of the effective potential $V$.

Let's consider the ground state of this system below the $\lambda$-temperature. In general, the ground state is given by solving the condition
\begin{eqnarray}
	\frac{\partial V}{\partial \varphi}
	 = -\mu\varphi^* + \lambda (\varphi^* \varphi) \varphi = 0. \label{eq:gsc}
\end{eqnarray}
This condition yields two solutions, 
\begin{eqnarray}
	\varphi = 0, \quad |\varphi| = \sqrt{\frac{\mu}{\lambda}} \label{eq:ground}. 
\end{eqnarray}
As shown in Fig.~\ref{fig:transition}, the trivial solution $\varphi = 0$ is not stable for $T < T_\lambda$, but another solution $|\varphi | = \sqrt{\mu/\lambda}$ is stable. Furthermore, because $|\varphi|^2$ is equal to the number density $n$ of superfluid, the following relation is satisfied below the $\lambda$-temperature,
\begin{eqnarray}
	|\varphi | = \sqrt{\frac{\mu}{\lambda}} = \sqrt{n} \equiv \sqrt{\frac{\rho_s}{m}}, 
	\label{eq:varphi}
\end{eqnarray}
where $\rho_s$ is the superfluid density and $m$ is the mass of a helium atom. From eq.~(\ref{eq:varphi}), the chemical potential at zero temperature $\mu (0)$ is given by 
\begin{eqnarray}
  \mu (0) = \frac{\lambda \rho_s (0)}{m} 
          = \frac{\lambda \rho_0}{m}, \label{eq:chemical}
\end{eqnarray}
where $\rho_0$ is the total density of liquid helium-4 at zero temperature, and we have used the relation $\rho_s (0) = \rho_0$. 

Before deriving the formula for $T_\lambda$, we should see the spontaneous symmetry breaking.
For $T > T_\lambda$, the effective potential has only one minimum at $\varphi = 0$, and then the global $U(1)$ symmetry is not broken. For $T < T_\lambda$, the field $\varphi$ rolls down from "hill-top" at $\varphi = 0$ to the bottom, and then the global $U(1)$ symmetry is spontaneously broken. This phenomenon is the $\lambda$-transition.

The derivation of the formula for $T_\lambda$ is very easy as follows. When $T=0$, the scalar field $\varphi$ is located at the bottom, and then it takes the nonzero value $|\varphi|=\sqrt{\mu (0)/\lambda}$. Hence the global $U(1)$  symmetry is spontaneously broken in this phase. As the temperature increases above the $\lambda$-temperature, the scalar field $\varphi$ is located at $\varphi = 0$, and the global $U(1)$ symmetry is recovered. Therefore, it is enough that we consider how energy is necessary to take the scalar field $\varphi$ from the bottom to "hill-top". From eq.~(\ref{eq:potential}), the depth of the potential $V_0$ is given by
\begin{eqnarray}
	V_0 \equiv -V(|\varphi| = \sqrt{\mu(0)/\lambda}) = \frac{\mu (0) n_0}{2}, \label{eq:depth}
\end{eqnarray}
where $n_0 \equiv \rho_0/m$ is the number density of liquid helium-4 at zero temperature. The depth of the effective potential represents the required energy density to recover the global $U(1)$ symmetry. Because the scalar field $\varphi$ has {\it two} degrees of freedom (its amplitude and phase), the $\lambda$-temperature is given by
\begin{eqnarray}
	V_0/n_0 = \frac{1}{2} k_B T_\lambda \times 2. \label{eq:energy-density}
\end{eqnarray}
\begin{figure}[t]
\includegraphics[width=5cm]{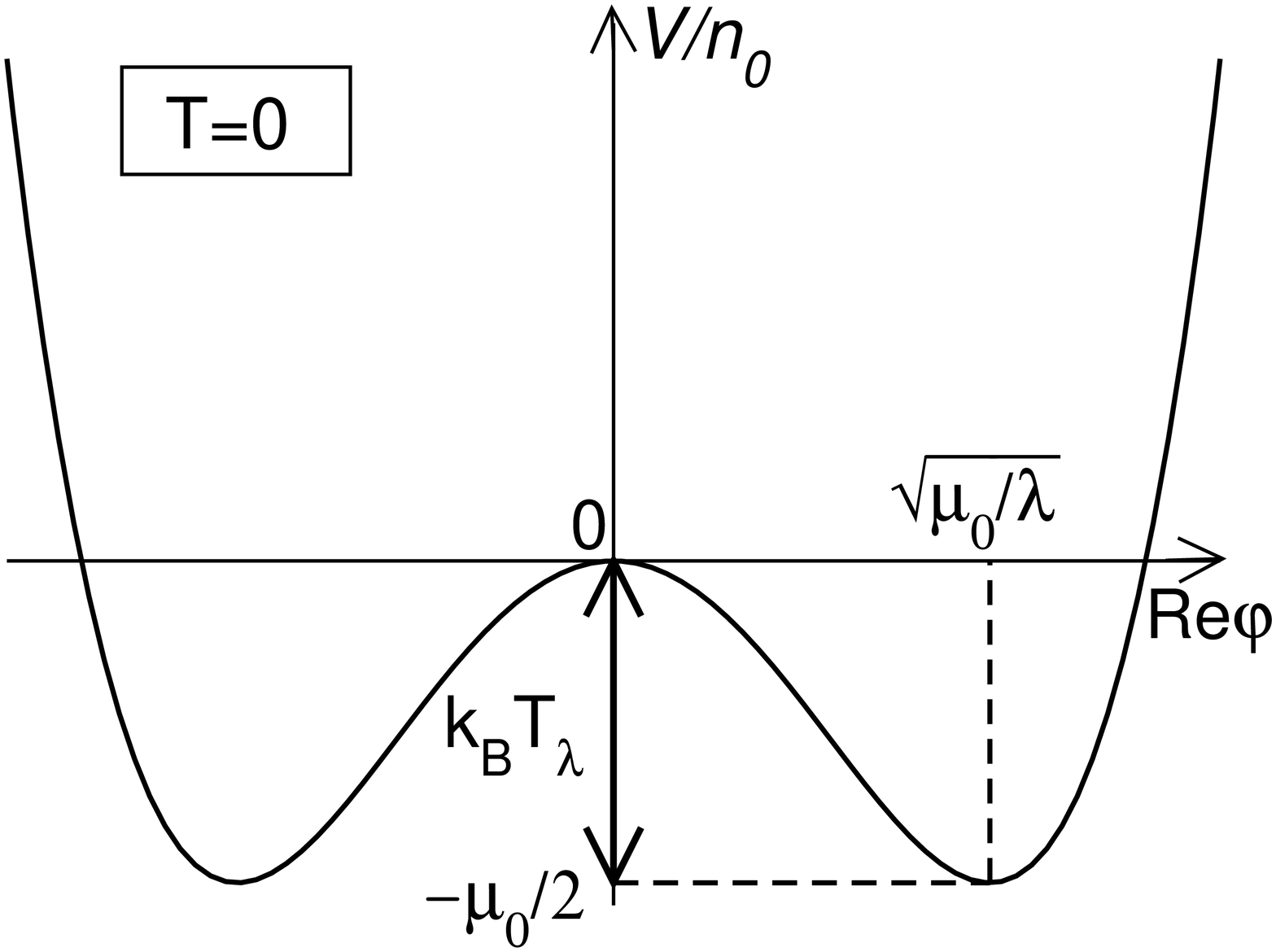}
\caption{\label{fig:absolute_zero} The potential per number density at zero temperature. The depth represents the $\lambda$-temperature.}
\end{figure}
Fig.~\ref{fig:absolute_zero} shows this situation. 
From eqs.~(\ref{eq:chemical}), (\ref{eq:depth}), (\ref{eq:energy-density}) and $\lambda = 2\pi \hbar^2 \sigma/m$, we obtain the following formula for the $\lambda$-temperature, 
\begin{eqnarray}
 T_\lambda = \frac{\rho_0 \pi \hbar^2 \sigma}{m^2 k_B}. \label{eq:lambda-2}
\end{eqnarray}

Using eq.~(\ref{eq:lambda-2}), one can calculate the $\lambda$-temperature, because the experimental values of $\rho_0$, $m$ and $\sigma$ are known. Substituting the following values into eq.~(\ref{eq:lambda-2}), 
\begin{eqnarray}
	\rho_0 & = & 0.1451 \ \mbox{g/cm$^3$}, \\
	m  & = & 6.6465 \times 10^{-24} \ \mbox{g}, \\
	\sigma & = & 2.639 \ \mbox{\AA},
\end{eqnarray}
one obtains the theoretical value
\begin{eqnarray}
	T_\lambda = 2.194 \ \mbox{K}.
\end{eqnarray}
This is very close to the experimental values of 2.1768 K, and the deviation between the theoretical and experimental values is less than 1 \%. Thus, spontaneous symmetry breaking is a key concept to understand the $\lambda$-transition.

\section{Roton energy gap}
In this section, we calculate the roton energy gap at zero temperature $\Delta (0)$ using the effective potential~(\ref{eq:potential}).\\

\noindent (i) {\it linear decomposition}

Because we are interested in the excitations near the ground state, it is convenient to define 
\begin{eqnarray} 
	\varphi (\vec{x},t) = \varphi_c + \eta (\vec{x},t) + i\chi (\vec{x},t), \quad \varphi_c \equiv \sqrt{\frac{\mu}{\lambda}}, \label{eq:linear}
\end{eqnarray} 
where $\eta$ and $\chi$ are {\it real} scalar fields. This decomposition is known as {\it linear decomposition}. Plugging (\ref{eq:linear}) into (\ref{eq:potential}), we obtain the effective potential in terms of $\eta$ and $\chi$
\begin{eqnarray}
	V = -\frac{\mu^2}{2\lambda} + 2\mu \eta^2 
		+ 2\sqrt{\lambda\mu} (\eta^2 + \chi^2) \eta 
		+ \frac{\lambda}{2}(\eta^2 + \chi^2)^2. \ \label{eq:potential-linear}
\end{eqnarray}

In eq.~(\ref{eq:potential-linear}), the first term (at zero temperature) determines the $\lambda$-temperature as shown in section 2, 
\begin{eqnarray}
	V_0 = \frac{\mu (0) ^2}{2\lambda} = k_B T_\lambda n_0. \label{eq:V_0}
\end{eqnarray}
Using the relation $\mu (0)/\lambda = n_0$ (see, eq.~(\ref{eq:varphi})), we obtain
\begin{eqnarray}
	\mu (0) = 2 k_B T_\lambda. \label{eq:chemical_0}
\end{eqnarray}
This is the direct relation between the chemical potential at zero temperature and the $\lambda$-temperature. 

\begin{figure}[t]
\includegraphics[width=7cm]{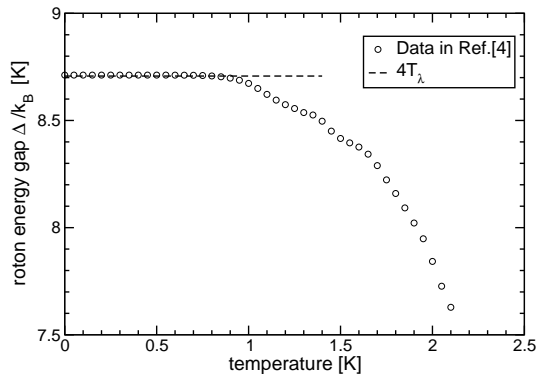}
\caption{\label{fig:roton-temp} The roton energy gap as a function of temperature. The theoretical value is in good agreement with data.}
\end{figure}

However, for the present purpose, the second term in eq.~(\ref{eq:potential-linear}) is more important, because the coefficient of $\eta^2$ term represents the potential energy of the real scalar field $\eta$. Therefore, the coefficient of $\eta^2$ should be identified with the energy gap of roton at zero temperature
\begin{eqnarray}
 \Delta (0) = 2 \mu (0). \label{eq:Delta_0}
\end{eqnarray}
From eqs.~(\ref{eq:chemical_0}) and~(\ref{eq:Delta_0}), the roton gap $\Delta (0)$ is given by
\begin{eqnarray}
	 \Delta (0) = 2 \mu (0) = 4k_B T_\lambda,
\end{eqnarray}
or
\begin{eqnarray}
	 \Delta (0)/k_B T_\lambda = 4. \label{eq:rotongap}
\end{eqnarray}
Substituting $T_\lambda = 2.1768$ K into eq.~(\ref{eq:rotongap}), one obtains
\begin{eqnarray}
	&& \Delta (0)/k_B = 8.707 \ \mbox{K} \quad (\mbox{theory}), \\
	&& \Delta (0)/k_B = 8.712 \ \mbox{K} \quad (\mbox{experiment}).
\end{eqnarray}
Thus, our relation (\ref{eq:rotongap}) is very precise. The deviation between the theoretical and experimental values is less than 0.1\%.
Fig.~\ref{fig:roton-temp} shows the roton energy gap as a function of temperature. From fig.~\ref{fig:roton-temp}, one can find that the relation $\Delta (0)/k_B T_\lambda = 4$ is in good agreement with the experimental values at low temperatures ($T < 0.8$ K).

Here we comment on the energy gap of phonon. Because the coefficient of $\chi^2$ term in eq.~(\ref{eq:potential-linear}) is zero, the energy gap of the scalar field $\chi$ is exactly zero. This means that phonon is gapless, and it is a consequence of the Nambu-Goldstone theorem.\\

\noindent (ii) {\it polar decomposition}

Although the linear decomposition is enough to calculate the roton energy gap, we consider another decomposition for completeness
\begin{eqnarray} 
	\varphi (\vec{x},t) = \left(\varphi_c + \sigma (\vec{x},t) \right) e^{i\pi(\vec{x},t)/\varphi_c}, \quad \varphi_c \equiv \sqrt{\frac{\mu}{\lambda}}, \label{eq:polar}
\end{eqnarray} 
where $\sigma$ and $\pi$ are {\it real} scalar fields. This is known as {\it polar decomposition}. Because the field $\pi$ is divided by $\varphi_c$, the fields $\sigma$ and $\pi$ have the same physical dimensions. In this decomposition, the field $\pi$ is the Nambu-Goldstone mode and $\sigma$ is the Higgs mode. 

Plugging (\ref{eq:polar}) into (\ref{eq:potential}), we obtain the effective potential in terms of $\sigma$
\begin{eqnarray}
	V = -\frac{\mu^2}{2\lambda} + 2\mu \sigma^2 + 2\sqrt{\lambda\mu} \sigma^3+ \frac{\lambda}{2}\sigma^4. \label{eq:polarpotential}
\end{eqnarray}
Thus, we find that the Nambu-Goldstone mode $\pi$ vanishes in the effective potential. The field $\pi$ appears only in the kinetic terms in the Lagrangian. Therefore, the Nambu-Goldstone mode or phonon is clearly gapless. 

The physical meaning of the 1st and 2nd terms in eq.~(\ref{eq:polarpotential}) is the same as that of linear decomposition. The 1st term determines $T_\lambda$, and the 2nd term determines $\Delta (0)$. Thus, the linear decomposition and the polar decomposition yield the same physical results; phonon is gapless and the roton gap is given by $\Delta (0)/k_B T_\lambda = 4$. 

\section{Curvature matrix}
In this section, we introduce the {\it curvature matrix}, which allows us to calculate the energy gap of elementary excitations more directly. 

First, let's consider the effective potential at $T=0$
\begin{eqnarray}
  V(0) 
  = -\mu (0)\varphi^*\varphi + \frac{\lambda}{2}(\varphi^*\varphi)^2. 
  \label{eq:potential-zero}
\end{eqnarray}
Here we introduce the $2 \times 2$ curvature matrix $C$
\begin{eqnarray} 
 C \equiv
 \begin{pmatrix}
  \frac{\partial^2 V(0)}{\partial \varphi^* \partial \varphi^*}
  \Big|_{\varphi = \varphi_{c}(0)} & 
  \frac{\partial^2 V(0)}{\partial \varphi^* \partial \varphi}
  \Big|_{\varphi = \varphi_{c}(0)} \\
  \frac{\partial^2 V(0)}{\partial \varphi^* \partial \varphi}
  \Big|_{\varphi = \varphi_{c}(0)} &
  \frac{\partial^2 V(0)}{\partial \varphi \partial \varphi}
  \Big|_{\varphi = \varphi_{c}(0)} 
 \end{pmatrix},
\end{eqnarray} 
where $\varphi_c (0)$ is the field value at $T=0$ (see, eq.~(\ref{eq:ground})), 
\begin{eqnarray}
 \varphi_c (0) \equiv \sqrt{\frac{\mu (0)}{\lambda}}.
\end{eqnarray}
Because the matrix $C$ represents the curvature at the bottom of the effective potential, two eigenvalues of the matrix $C$ should be identified with the energy gap of elementary excitations at zero temperature. 

From eq.~(\ref{eq:potential-zero}), the curvature matrix $C$ is given by
\begin{eqnarray} 
 C = \mu (0)
 \begin{pmatrix}
  1 & 1 \\ 1 & 1
 \end{pmatrix}
   = 2 k_B T_\lambda
 \begin{pmatrix}
  1 & 1 \\ 1 & 1
 \end{pmatrix}.
\end{eqnarray} 
Therefore, two eigenvalues of the matrix $C$ are 
\begin{eqnarray}
 0, \quad 4 k_B T_\lambda.
\end{eqnarray}
These eigenvalues represent the energy gap of elementary excitations;
\begin{eqnarray}
 \Delta_{\mbox{{\tiny phonon}}} & = & 0  \quad (\mbox{Nambu-Goldstone mode}) \\
 \Delta_{\mbox{{\tiny roton}}}  & = & 4 k_B T_\lambda \quad (\mbox{Higgs mode}).
\end{eqnarray}
Thus, using the curvature matrix, one can calculate the energy gap of elementary excitations very easily.

\section{Phonon}
Finally, we comment on an open problem about the sound speed of phonon. 

\begin{figure}[t]
\includegraphics[width=7cm]{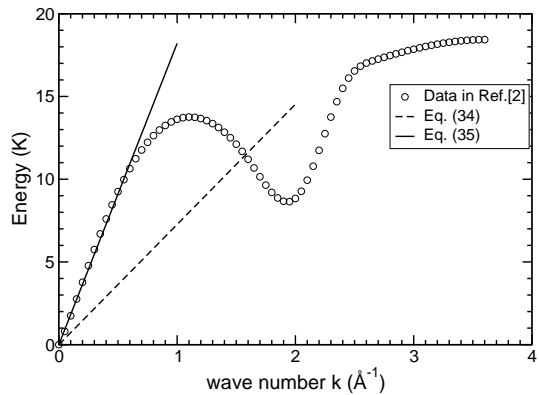}
\caption{\label{fig:dispersion} Phonon-roton dispersion curve.}
\end{figure}

According to the standard Bogoliubov theory~\cite{Bogoliubov:1947}, it is well known that the sound speed of phonon $c$ is given by 
\begin{eqnarray}
 c = \sqrt{\frac{\mu (0)}{m_4}}, \label{eq:soundspeed}
\end{eqnarray}
where $m_4$ is the mass of a helium-4 atom and $\mu (0)$ is the chemical potential at $T=0$. Although the original Bogoliubov theory cannot calculate the sound speed {\it quantitatively}, we can calculate it because we have derived the relation (\ref{eq:chemical_0}). From eqs.~(\ref{eq:chemical_0}) and (\ref{eq:soundspeed}), one obtains 
\begin{eqnarray}
 c = \sqrt{\frac{\mu (0)}{m_4}} = \sqrt{\frac{2 k_B T_\lambda}{m_4}}
   = 95 \ \mbox{m/s},
   \label{eq:soundspeed_m4}  
\end{eqnarray}
where we have used $m_4 = 6.64645 \times 10^{-24}$ g. This value is clearly inconsistent with the experimental value of 238 m/s. Thus, although the original Bogoliubov theory explains the linear dispersion of phonon ($\varepsilon = cp = c\hbar k$), it fails to explain the sound speed quantitatively.

Here we propose a possible expression of sound speed, which is consistent with the experimental value. In eq.~(\ref{eq:soundspeed}) (of Bogoliubov theory), if we replace $m_4$ by the effective roton mass $m_* = m_4/2\pi = 0.16 \ m_4$, one obtains
\begin{eqnarray}
 c = \sqrt{\frac{\mu (0)}{m_*}} = \sqrt{\frac{2 k_B T_\lambda}{m_*}}
   = 238 \ \mbox{m/s}.
   \label{eq:soundspeed_m*}  
\end{eqnarray}
Surprisingly, this is completely consistent with the experimental value as shown in fig.~\ref{fig:dispersion}, but its physical origin still remains to be solved. This indicates that the roton effective mass (or equivalently, the roton kinetic energy) should be considered in order to calculate the sound speed of phonon. However, at present, the theoretical explanation for eq.~(\ref{eq:soundspeed_m*}) is not known. We hope that this open problem will be discussed in the community.

\section{conclusion}
In conclusion, we have calculated the roton energy gap at zero temperature using an approach based on the spontaneous symmetry breaking. The relation that we have derived, $\Delta (0)/k_B T_\lambda = 4$, is in good agreement with the experimental value. We hope that this work contributes to the progress of low temperature physics.

%\acknowledgements

\end{document}